\documentclass[11pt]{article}

\usepackage[parfill]{parskip}
\usepackage{xcolor}
\usepackage{natbib}
\usepackage{bm}
\usepackage{amsmath}
\usepackage{graphicx}
\usepackage{rotating}
\usepackage{amsfonts}

\usepackage[margin=2.5cm]{geometry}
\usepackage{hyperref} 

\title{Assessing course difficulty and the effect of weather in amateur cross country running races}
\author{Kevin J. Wilson$^{1}$ \& Nina Wilson$^{2}$ \\
	$^1$School of Mathematics, Statistics \& Physics, Newcastle University, UK \\
	$^2$Biostatistics Research Group, Population Health Sciences Institute, Newcastle University, UK}
\date{}

\begin{document}
	
	\maketitle		
		
\begin{abstract}
Cross country running races are different to track and road races in that the courses are not typically accurately measured and the condition of the course can have a strong effect on the finish times of the participants. In this paper we investigate these effects by modelling the finish times of all participants in 28 cross country running races over 5 seasons in the North East of England. We model the natural logarithm of the finish times using linear mixed effects models for both the senior men's and senior women's races. We investigate the effects of weather and underfoot conditions using windspeed and rainfall as covariates, fit distance as a covariate, and investigate the effect of time via the season of the race, in particular investigating any evidence of a pre- to post-Covid effect. We use random athlete effects to model the participant to participant variability and identify the most difficult courses using random course effects. The statistical inference is Bayesian. We assess model adequacy by comparing samples from the posterior predictive distribution of finish times to the observed distribution of finish times in each race. We find strong differences between the difficulty of the courses, effects of rainfall in the month of the race and the previous month to increase finish times and an effect of increasing distance increasing finish times. We find no evidence that windspeed affects finish times.
\end{abstract}

{\bf Keywords:} linear mixed effects models, Bayesian inference, endurance running, race times

\section{Introduction}

Cross country racing is a form of running which takes place in the winter months over grass, fields, and other soft surfaces. It differs from road and track racing in that distances tend not to be exactly measured, and finish times are not directly comparable from year to year on the same course due to differing weather and underfoot conditions. Cross country racing is an elite discipline which attracts many of the top endurance runners in the world, with National, Continental and World Championships taking place each year. In addition, it is a popular amateur sport, with junior and local leagues operating within individual countries. For example, in England there are 39 counties which organise cross country championships each year.

In this paper, we consider one amateur cross country league, the North East Harrier League (NEHL) in England, which organises a series of 6 cross country races each season between September and March. Using the results of all races staged by the league between the 2017/18 and 2022/23 seasons, we investigate the effects of different courses, different seasons (to investigate a possible time trend), the weather and the underfoot conditions. We also take into account the fact that the same individuals take part in multiple races within and between seasons.

There has been a body of work establishing the effects of weather and environmental conditions on running performances, particularly in marathons \citep{Zha92,Ely07,Vim10,Kne19,Ber19,Wei22}. While most focus on effects of high temperatures and humidity, not relevant to cross country running in the North East of England, \cite{Kne19} investigated performances in the Boston marathon and found significant negative effects of windspeed and rainfall. Similarly, \cite{Vim10} found a negative effect of rainfall on the marathon performance of male runners, although they attributed this effect to the negative correlation between rainfall and air temperature. \cite{Ber19} found effects of distance and elevation on the pace of runners over different distances in road races.

Researchers have developed prediction models for endurance running events, typically either for track or road races \citep{Keo19,Alv20}. In a review article, \cite{Alv20} found that prediction models considered to date were mainly focused on laboratory testing variables such as VO$_2$max, training variables such as training pace and training load and anthropometric variables such as fat mass and skinfolds. \cite{Keo19} performed a review of 114 prediction models for marathon times. They found that while there were some models with high $R^2$ values based on training, laboratory and anthropometric variables, important variables such as course gradient, sex, and weather conditions were typically not included, limiting the usefulness of such models.

More statistical investigations into endurance running and racing have examined the impact the development and widespread adoption of super-shoes has had on endurance racing times for elite athletes via extreme value models \citep{Ard22} and developed a predictive framework to estimate the probability a runner will reach the next checkpoint and expected passage time at the checkpoint in ultra-distance trail races \citep{Fog21}. \cite{Sti23} considered the prediction of race times for elite runners in track events in the presence of missing data and proposed a latent class matrix-variate state space model, demonstrating that taking into account missing data patterns improves predictive performance.

In this paper we examine the difficulty of each course in the NEHL, based on all of the senior (individuals at least 18 years old) finish times over a 5 season period. We also look to see if there is evidence of the effect of weather and underfoot conditions on finish times, and investigate whether there is a time-effect, in particular looking to see if there is any evidence of a pre- to post-Covid difference. We consider men's and women's results separately, both to identify any sex-based differences and because men and women run different distances in the NEHL - with the longer men's races requiring more endurance than the shorter women's races. To investigate these questions we

\begin{enumerate}
	\item Perform exploratory data analysis on all of the data for men and women in the five seasons of the NEHL considered.
	\item Fit linear mixed effects models via Bayesian inference to the men's and women's data separately, investigating both log finish times and log pace as response variables, and including a distance covariate to allow for linear effects of distance on (log) finish times.
	\item Investigate summaries of the posterior predictive distributions of finish times to assess model adequacy, and validate the inference from the linear mixed effects models.
\end{enumerate}

The rest of the paper is structured as follows. In Section \ref{sec:dataexp} we provide a description of the data used in the analyses in the paper and then undertake an initial exploratory analysis. In Section \ref{sec:model} we describe the linear mixed effects models, their implementation via Bayesian inference and the results from the modelling. In Section \ref{sec:fit} we assess model adequacy via an assessment of the posterior predictive distributions of finish times. In Section \ref{sec:discuss} we discuss the results and conclude the paper.

\section{Data and exploratory data analysis}
\label{sec:dataexp}

\subsection{Data}

We consider all senior men's and women's results from the NEHL between the 2017/18 season and the 2022/23 season. The data are available publicly via the league's website \url{www.harrierleague.com}. In all, there are 14,067 men's finish times and 10,515 women's times across 28 races on 8 different courses, made up of 2668 unique male and 2116 unique female athletes. A breakdown of the number of finishers by age group across all races is provided in Figure \ref{fig:age} in the Appendix. The courses visited during the 5 season period are named Alnwick, Aykley Heads, Druridge Bay, Gosforth, Herrington, Lambton, Thornley and Wrekenton. Note that there are two races missing from the 2019/20 season, which were cancelled due to the Covid-19 pandemic, and there were no races in the 2020/21 season. 

Alongside the finish time in the dataset we also have an identification number for each athlete, the course used, the NEHL season of the race, the distance of the race, the windspeed for the race and the month and year the race took place. The distances for the men's races were taken via GPS using a Garmin Forerunner 235 watch. Each race is a lapped course, where men run three laps (approximately 10km) and women two laps (approximately 6.5km). The women's distances were taken to be $2/3$ of the men's distances. The windspeeds were taken to be the windspeed recorded at the Durham weather station, within the NEHL area, at 2pm on the day of the race, taken from \url{www.visualcrossing.com}. In addition, we have data on the recorded monthly rainfall at the Durham weather station in each month of the period, taken from \url{https://www.metoffice.gov.uk/pub/data/weather/uk/climate/stationdata/durhamdata.txt}. This will be our proxy for the underfoot conditions in the races.

There are no missing data in the dataset. The finish times are accurately recorded using chip timing. There will be some measurement error in the race distances due to the nature of wearable GPS devices, and the windspeeds and rainfall as a result of them being measured at a central weather station rather than on the individual courses themselves. We will not model these measurement errors specifically, although future work could consider this.

\subsection{Exploratory data analysis}

\begin{figure}[ht]
\centering
\includegraphics[width=0.9\linewidth]{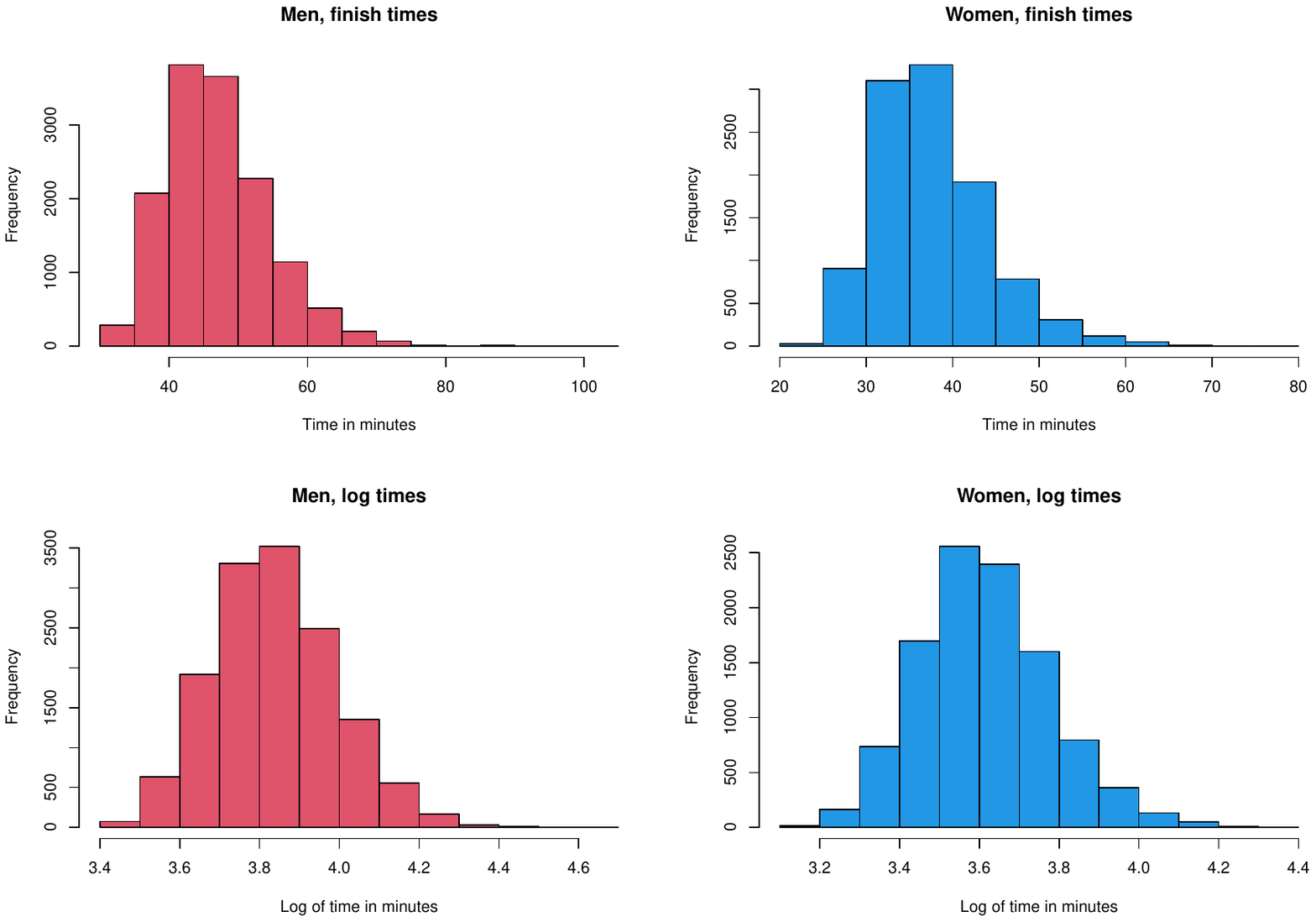}
\caption{Top: the distribution of finish times across all 28 races for men (left, red) and women (right, blue). Bottom: the distribution of the natural logarithm of finish times for men (left, red) and women (right, blue).}
\label{fig:hist}
\end{figure}

Initially we consider the overall distributions of finish times for men and women. These are given in the top row of Figure \ref{fig:hist}. We see that in both cases the finish times have a unimodal distribution with a positive skew. In the bottom row of Figure \ref{fig:hist} we provide histograms of the natural logarithm of the finish times for mean and women. We see that these are much more symmetrical around the centre of the distribution. Note that in both sets of plots the women's times are typically faster than the men's times due the shorter distance of the women's event in the NEHL.

\begin{figure}[ht]
	\centering
	\includegraphics[width=0.9\linewidth]{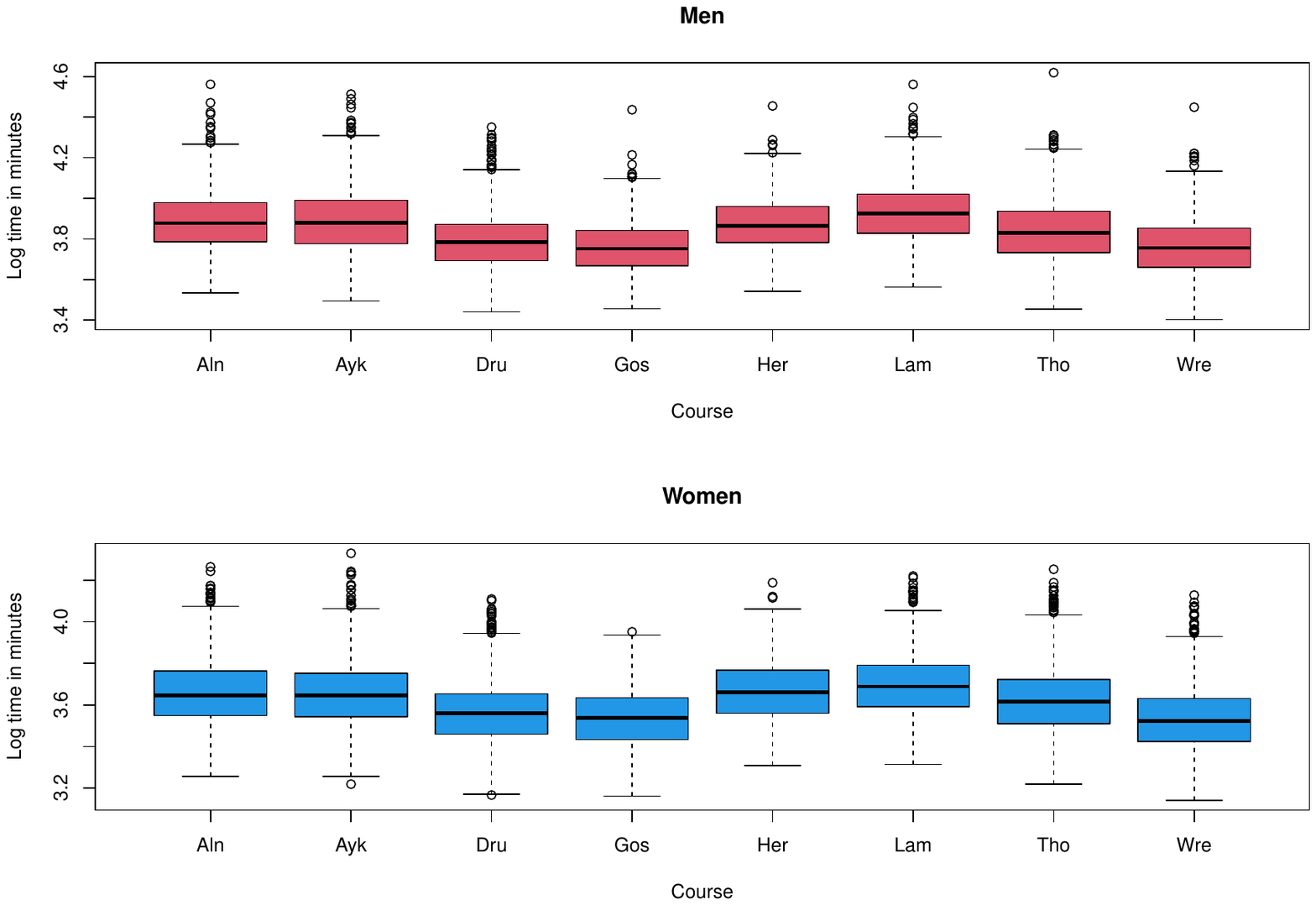}
	\caption{Boxplots of the distributions of the natural logarithm of finish times on each course for men (top) and women (bottom). The abbreviations of the course names stand for: Aln -Alnwick, Ayk - Aykley Heads, Dru - Druridge Bay, Gos - Gosforth, Herr - Herrington, Lam - Lambton, Tho - Thornley, Wre - Wrekenton.}
	\label{fig:box}
\end{figure}

In Figure \ref{fig:box} we provide boxplots of the distribution of the natural logarithm of the finish times for each course for men (top) and women (bottom). We see the same patterns of finish times across men and women - Druridge Bay, Gosforth and Wrekenton have the lowest median finish times and Lambton the highest. We see, however, substantial overlap between the distributions of finish times between all of the 8 different courses. There are outliers in the upper tail of each distribution indicating that there is still some slight positive skew present, representing runners who are substantially slower than a typical runner.

\begin{figure}[ht]
	\centering
	\includegraphics[width=0.9\linewidth]{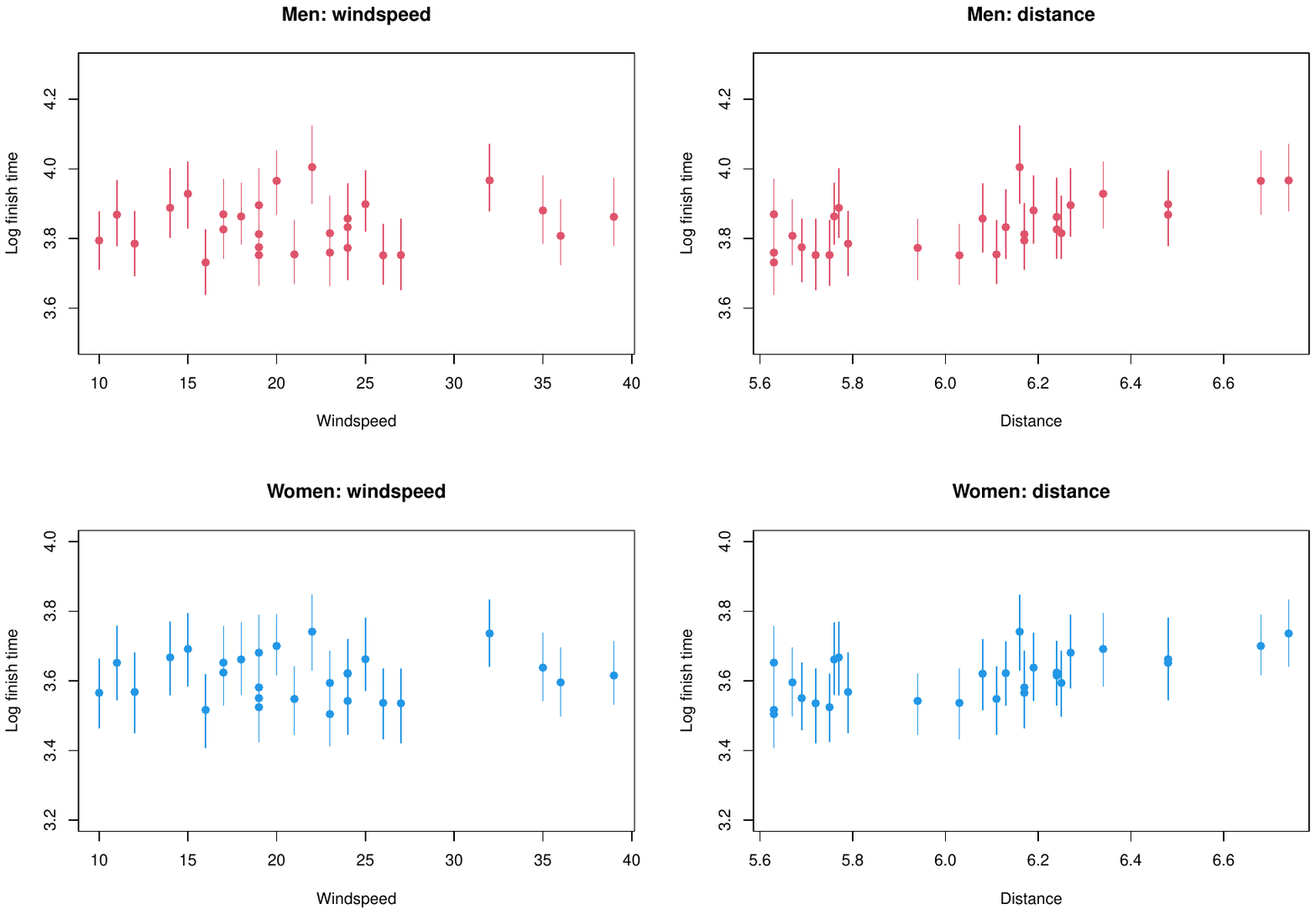}
	\caption{The median and inter-quartile range of the log finish times for each race plotted against the windspeed (left) and distance (right) for men (top, red) and women (bottom, blue).}
	\label{fig:cov}
\end{figure}

In Figure \ref{fig:cov} we plot the windspeed and distance against the log finish times for both men and women. In each case the plots are made up of the median log finish time (points) and interquartile range (vertical bars) for each race in each season on the y-axis, and the windspeed or distance on the x-axis. We see that for both men (red, top) and women (blue, bottom) there is no clear relationship between the windspeed and the distribution of finish times based on these data. However, for both men and women there appears to be a positive linear relationship between the distance of the race and the finish time on the log scale - the further the race, the slower the time on average. The distributions of finish times for each race show very similar patterns for men and women, although there is a shift in the y-axis values.  

In the Appendix we also include a set of plots (Figure \ref{fig:rain}) of the rainfall in the month of each race and the rainfall in the previous month, against the median and inter-quartile ranges of log finish times. While we see similar patterns between men and women, we see no clear relationship between rainfall and finish times.

\section{Statistical modelling}
\label{sec:model}

\subsection{The linear mixed effects model}

Our aim is to understand the relative difficulty of each of the eight courses used by the NEHL. We would also like to understand the effect of the covariates windspeed, distance and rainfall on the finish times of runners in NEHL races, and to see if we can detect any season to season variation. Finally, the same runners take part in multiple races across the five seasons, and so we need to make provision for the lack of independence of observations on the same individuals.

Thus we propose the following linear mixed effects model. The response $Y_{ijk}$ for individual $i$ on course $j$ in season $k$ is defined to be the natural logarithm of the finish time $T_{ijk}$, that is $Y_{ijk}=\log(T_{ijk})$. We assume $Y_{ijk}\mid\mu_{ijk},\tau\sim\textrm{N}(\mu_{ijk},1/\tau)$, where $\tau$ is the observation precision and the mean $\mu_{ijk}$ takes the form of a linear predictor
\begin{eqnarray*}
	\mu_{ijk} & = & \lambda + \alpha_i + \beta_j + \delta_k + \gamma(D_{jk} - \bar{d}) + \hspace{3pt} \lambda (W_{jk} - \bar{w}) + \rho_m R_{m(jk)} + \rho_{m-1}R_{m(jk)-1}, 
\end{eqnarray*} 
where $\lambda$ is an overall mean, $\alpha_i\sim\textrm{N}(0,1/\tau_\alpha)$ is a random athlete effect, $\beta_j\sim\textrm{N}(0,1/\tau_\beta)$ is a random course effect, $\delta_k\sim\textrm{N}(0,1/\tau_\delta)$ is a random season effect, $\gamma$ is a fixed effect of distance, $\lambda$ is a fixed windspeed effect, $\rho_m$ is a fixed effect of rainfall in the current month and $\rho_{m-1}$ is a fixed effect of rainfall in the previous month. The covariates $D_{jk}$ and $W_{jk}$ are the distance and windspeed of course $j$ in season $k$. $R_{m(jk)}$ represents the rainfall in month the race at course $j$ in season $k$. The terms $\bar{d}$ and $\bar{w}$ are chosen values representing average distance and average windspeed, to centre the covariates.

That is, we treat the athlete, course and season as random effects centred on zero {\em a priori}, consider the linear effects of distance and windspeed on the log scale and include fixed effects of rainfall in both the current and previous month, to represent that underfoot conditions are a result of the weather conditions over a number of weeks leading up to the race. By including individual terms for each month, rather than summing them, we hope to understand the relative contribution of each.

Inference is Bayesian and so a prior distribution is required. In practice, we assign independent marginal prior distributions to each model parameter except the rainfall effects, which are correlated in the prior. We give each fixed effect independent Normal prior distributions
\begin{eqnarray*}
	&\gamma\sim\textrm{N}(m_\gamma,v_\gamma),~\lambda\sim\textrm{N}(m_\lambda,v_\lambda), \\
	&\rho_m\sim\textrm{N}(m_{\rho_m},v_{\rho_m}),~\rho_{m-1}\sim\textrm{N}(\phi m_{\rho_{m}},v_{\rho_{m-1}}),
\end{eqnarray*}
where we choose relatively large prior variances in each case. The rainfall effects are correlation through their means, which both depend on $m_{\rho_m}$, the mean effect of rainfall in the current month, which is given a Normal prior distribution with a mean of zero. The effect of rainfall in the previous month has mean $\phi m_{\rho_m}$, where $\phi<1$ would represent a stronger effect of rainfall in the current month than the previous month. We give $\phi$ a gamma prior distribution, $\phi\sim\textrm{Gamma}(a_\phi,b_\phi)$, where $a_\phi\leq b_\phi$. 

The precisions from the random effects are given independent gamma priors, to provide some conjugacy and speed up inference. Specifically,
\begin{eqnarray*}
	&\tau_\alpha\sim\textrm{Gamma}(a_\alpha,b_\alpha), ~ \tau_\beta\sim\textrm{Gamma}(a_\beta,b_\beta) ~\tau_\delta\sim\textrm{Gamma}(a_\delta,b_\delta)
\end{eqnarray*}
where the hyper-parameters are to be chosen. In order for each of our parameters to be identifiable in the inference, we require corner constraints. That is, we constrain $\alpha_1=\beta_1=\delta_1=0.$ Thus all other random effects are interpreted relative to these values.

Inference is conducted using Markov Chain Monte Carlo (MCMC) methods using the \verb|rjags| package in R \citep{Plu22}, which implements JAGS software. Specifically, JAGS uses a combination of Gibbs sampling, Metropolis-Hastings sampling and slice sampling steps to sample from the posterior distribution. In the analyses that follow we discard an initial 10,000 samples as burn-in, before running a further 1,000,000 iterations and thinning by 100, to obtain 10,000 approximately independent samples from the posterior distribution. Convergence and mixing is checked using traceplots and effective sample sizes. All analyses run in minutes on a standard desktop or laptop computer. Hyper-parameters are chosen to give relatively diffuse prior distributions.

In the results that follow in the next section we report a slightly simpler version of the model, with linear predictor 
\begin{eqnarray*}
	\mu_{ijk} & = & \lambda + \alpha_i + \beta_j + \delta_k + \gamma(D_{jk} - \bar{d}) + \rho_m R_{m(jk)} + \rho_{m-1}R_{m(jk)-1}, 
\end{eqnarray*}
that is, with no windspeed effect. This is because all models that we have fitted show no discernible effect of windspeed, with posterior distributions for windspeed centred around zero. See Table \ref{tab:wind} in the Appendix for posterior summaries from this model.

We also fitted the model with a slightly different response variable, using log pace $Y_{ijk}=\log(T_{ijk}/D_{jk})$ instead of log finish time, to understand if our course effects were being driven by distance in spite of the distance covariate in the model. The results were qualitatively the same, and so we are confident that this is not the case. A summary of these results are provided in Table \ref{tab:pace} in the Appendix. 

\subsection{Results}

\begin{figure}[ht]
	\centering
	\includegraphics[width=0.495\linewidth]{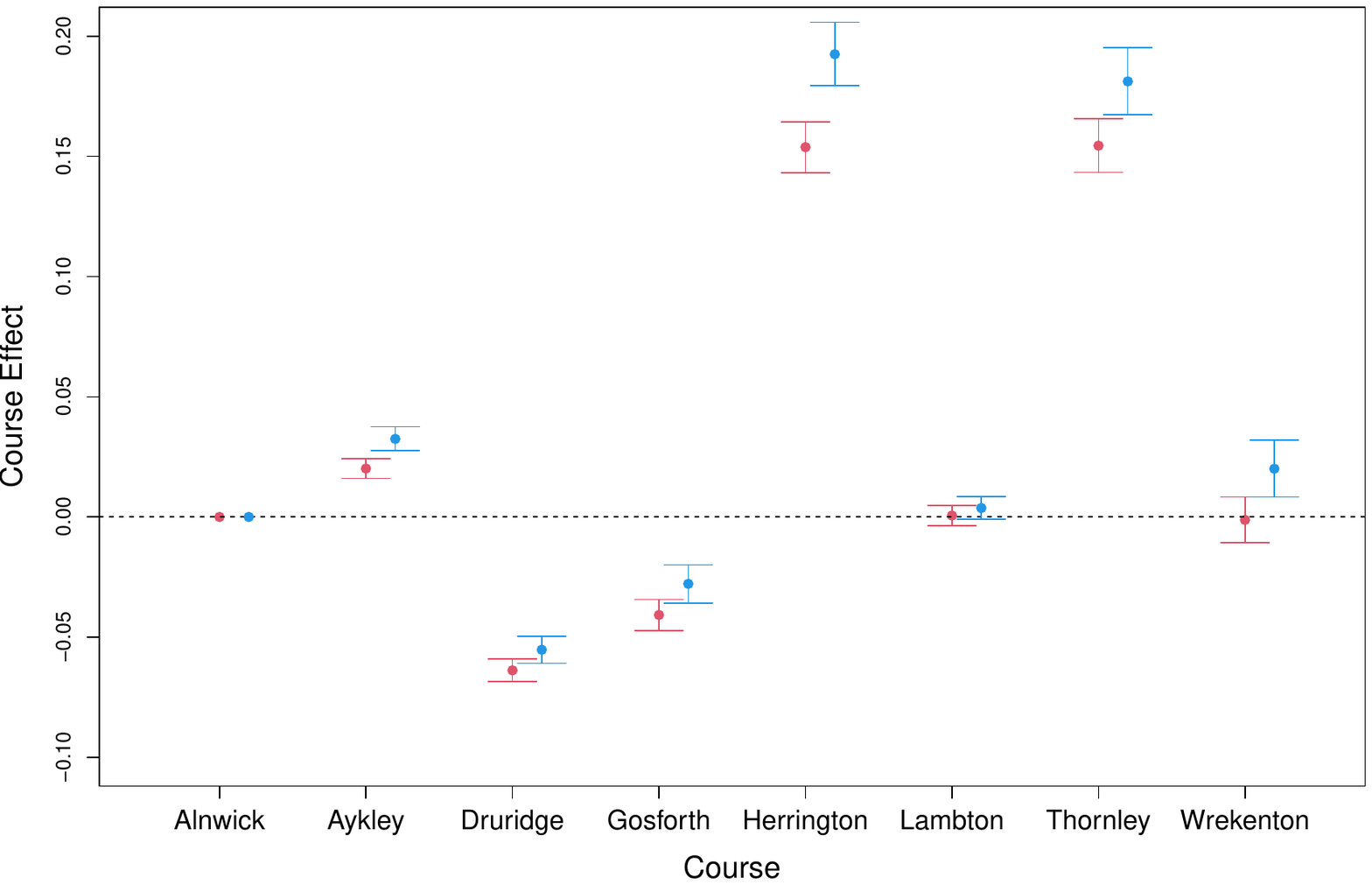}
	\includegraphics[width=0.495\linewidth]{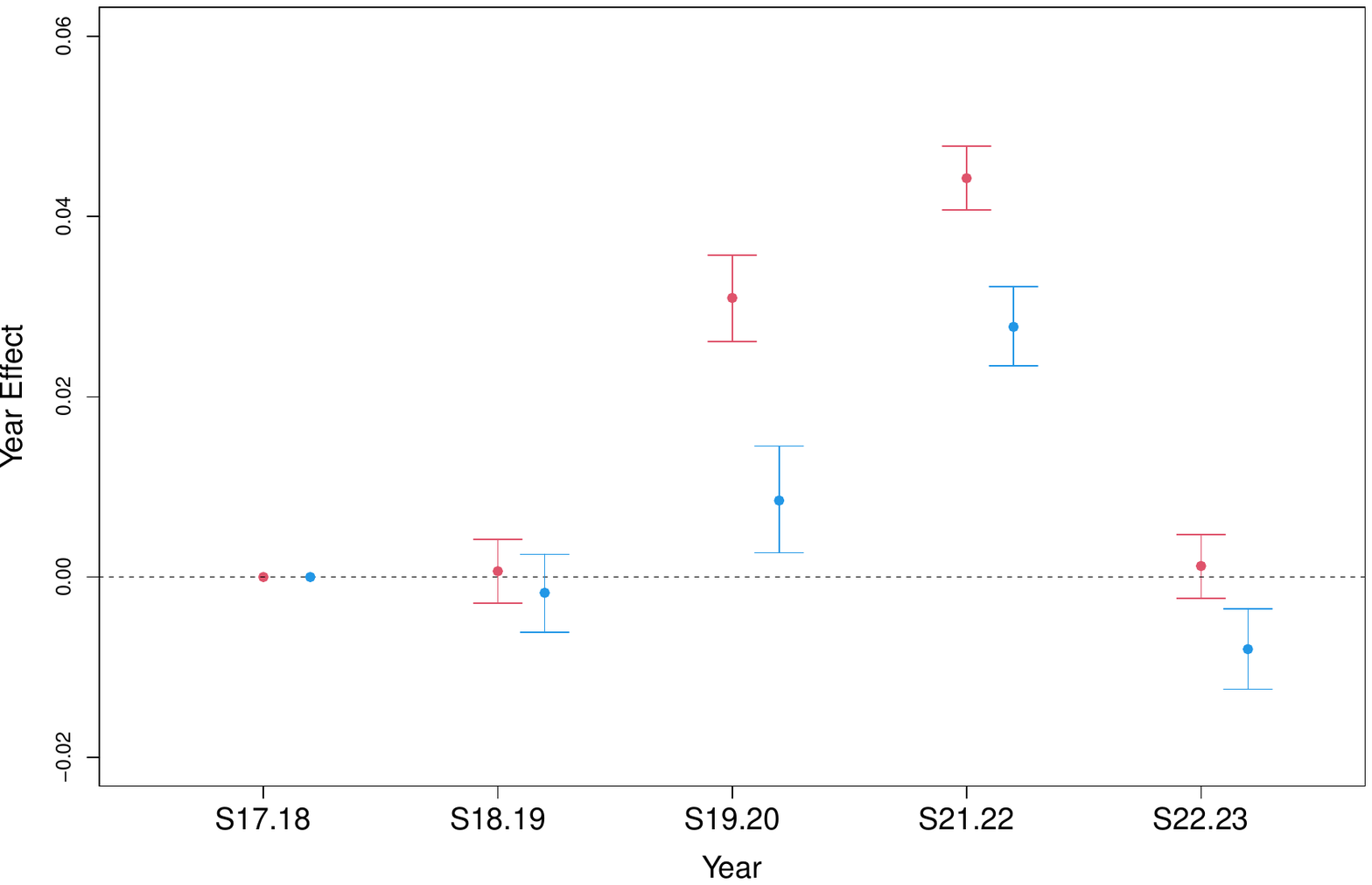}
	\caption{Left: the posterior mean and 95\% posterior symmetric intervals for the course effects for men (red) and women (blue). Right: the posterior mean and 95\% posterior symmetric intervals for the season effects for men (red) and women (blue).}
	\label{fig:course}
\end{figure}

First we consider the course effects. The left hand side of Figure \ref{fig:course} provides the posterior mean and 95\% posterior symmetric intervals for each of the courses for both men and women. We note that all effects are compared to Alnwick. We see that for men the most difficult courses are Herrington and Thornley, with the easiest courses being Druridge Bay and Gosforth. The ordering is almost identical for women, with relatively similar effects observed to the men, typically slightly higher. In general, we see much stronger differences between courses than in the exploratory analysis, now that the individual athlete effects and covariate effects have been removed. We provide the individual athlete effects in Figure \ref{fig:ath} in the Appendix. We see a large spread in athlete abilities in both the men's and women's races, with more evidence of a small number of very slow and very fast runners, compared to the rest, in the women's race.

We provide a plot of the posterior season effects for men (red) and women (blue) in the form of the posterior means and 95\% posterior symmetric intervals in right hand side of Figure \ref{fig:course}, for each of the five seasons in the data. We see no evidence of a systematic difference pre- and post-Covid, with the effects for both women and men being similar in Seasons 17/18, 18/19 and 22/23 and being positive in Seasons 19/20 and 21/22, representing slightly slower finish times on average. While we may have surmised that this may be due to either a lack of fitness in runners, or a different cohort made up of a higher percentage of new runners in 21/22, this would not explain the positive season effect in 19/20 prior to the pandemic. Overall, we conclude that there are no obvious temporal trends present.

\begin{figure}[h]
	\centering
	\includegraphics[width=0.9\linewidth]{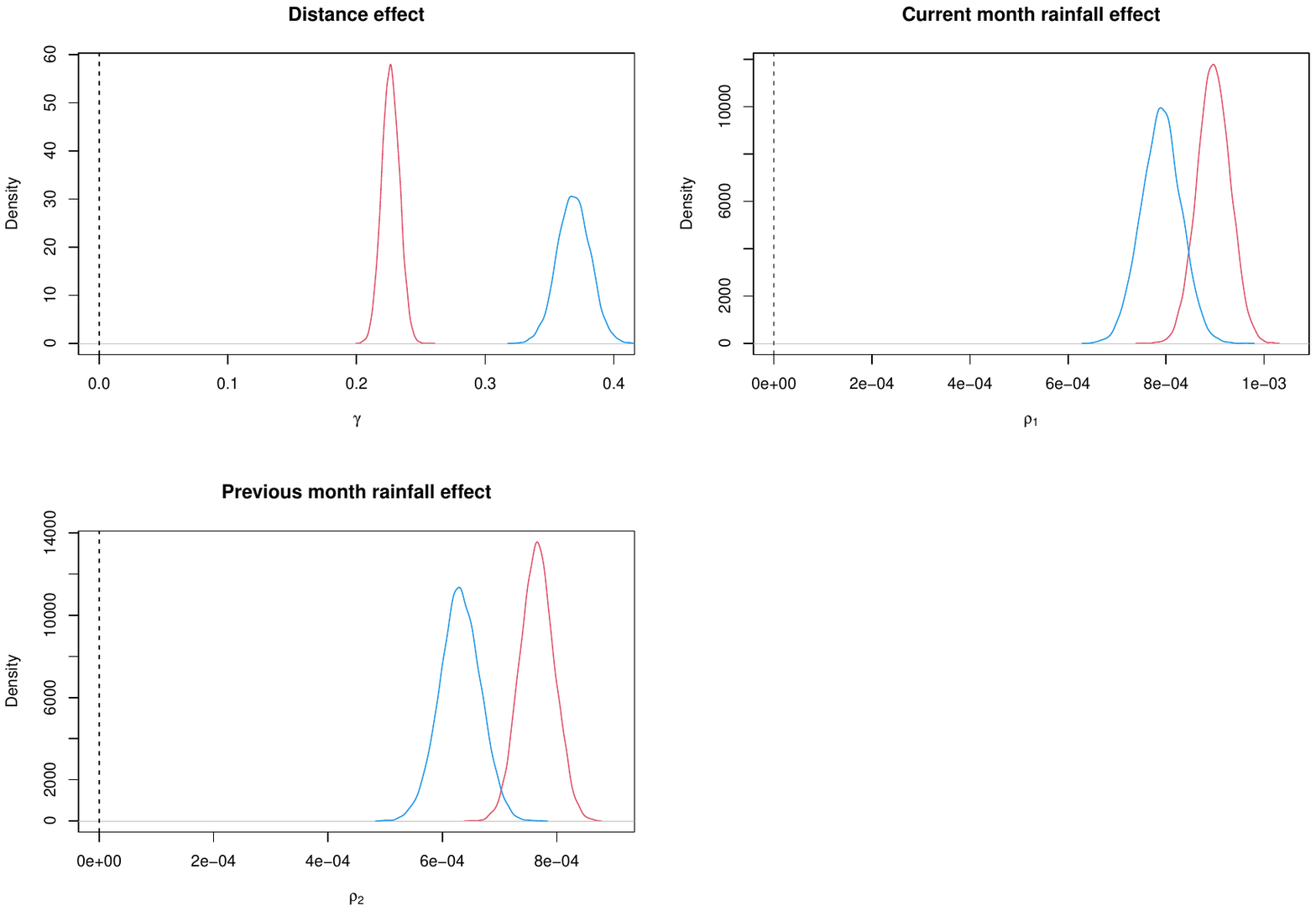}
	\caption{The posterior densities for the covariates distance (top left), rainfall in the current month (top right) and rainfall in the previous month (bottom left) for men (red) and women (blue).}
	\label{fig:post}
\end{figure}

We can investigate the effects of the covariates in the model. We plot their posterior densities in Figure \ref{fig:post} for men (red) and women (blue). We see that for rainfall in the current and previous month the effect is stronger for men than for women.The opposite is the case for distance. In each case the qualitative interpretation of the effect is the same for men and women.

\begin{itemize}
	\item The distance effect is positive, so as distance increases, so does the log finish time. Based on the posterior means of the distance effects, a 0.1 mile increase in distance would lead to an increase in finish time of 1 minute 5 seconds for a man running a 47 minute race and 1 minute 26 seconds for a woman running a 38 minute race.
	\item The rainfall effect in the current month is positive, so as rainfall increases so do log finish times. Based on the posterior means of the current rainfall effects, a 10mm increase in the current month's rainfall would lead to an increase in finish time of 25 seconds for a man running a 47 minute race and 18 seconds for a woman running a 38 minute race.
	\item The rainfall effect in the previous month is positive, so as rainfall increases so do log finish times. Based on the posterior means of the previous rainfall effects, a 10mm increase in the previous month's rainfall would lead to an increase in finish time of 22 seconds for a man running a 47 minute race and 14 seconds for a woman running a 38 minute race.
\end{itemize}

\subsection{Model checking}
\label{sec:fit}

We assess the fit of our models to the data in this section using posterior predictive distributions. We compare the distribution of (log) finish times based on a sample from the posterior predictive distributions of the (log) linear mixed effects models to the observed distributions of finish times for all of the races in the 17/18 to 22/23 seasons of the NEHL. We do so via the histograms of the full distributions of finish times and summary statistics from these distributions. This will allow us to see if our models are adequately capturing the variation in finish times in the data, validating the inferences made in the previous section.

We sample results from each of the 28 races in the five NEHL seasons considered, from the posterior predictive distribution, utilising the information on the course, athletes, season and covariates. We overlay the resulting distribution of log finish times for each race (blue) and the observed distribution of log finish times (red) in Figures \ref{fig:predmen} and \ref{fig:predwomen} for men and women respectively in the Appendix.

We see that for both the men's and women's races there are no obvious discrepancies between the sample of log finish times from the posterior predictive distributions and the observed log finish times. They are remarkably similar. We investigate this further by considering summaries from these two distributions, reported in Table \ref{tab:pred}, for three randomly selected races: Aykley Heads in 2017, Druridge Bay in 2022 and Wrekenton in 2021.

\begin{table}
	\centering
	\begin{tabular}{|cc|c|ccccc|} \hline
		\multicolumn{8}{|l|}{{\bf Men}} \\ \hline
		Course & Year & Obs/Pred & Min & LQ & Med & UQ & Max \\ \hline
		Aykley Heads & 2017 & Obs & 34.47 & 42.06 & 46.02 & 51.32 & 77.20 \\ 
		& & Pred & 30.64 & 42.67 & 46.40 & 52.17 & 83.83  \\ \hline
		Druridge Bay & 2022 & Obs & 31.58 & 39.72 & 43.42 & 46.98 & 74.68 \\
		& & Pred & 30.32 & 39.66 & 43.42 & 48.04 & 72.22 \\ \hline
		Wrekenton & 2021 & Obs & 31.95 & 39.99 & 43.98 & 48.18 & 85.55 \\
		& & Pred & 30.59 & 39.73 & 43.30 & 47.61 & 64.77  \\ \hline 
		\multicolumn{8}{|l|}{{\bf Women}} \\ \hline
		Course & Year & Obs/Pred & Min & LQ & Med & UQ & Max \\ \hline
		Aykley Heads & 2017 & Obs & 27.02 & 34.12 & 37.05 & 40.91 & 60.00 \\ 
		& & Pred & 27.32 & 34.38 & 37.78 & 41.73 & 61.86 \\ \hline
		Druridge Bay & 2022 & Obs & 31.58 & 39.72 & 43.42 & 46.98 & 74.68 \\
		& & Pred & 21.99 & 31.01 & 34.50 & 37.86 & 57.44 \\ \hline
		Wrekenton & 2021 & Obs & 23.82 & 31.38 & 34.72 & 38.85 & 62.10  \\
		& & Pred & 23.21 & 31.42 & 35.11 & 39.05 & 58.23  \\ \hline 
	\end{tabular}
	\caption{Summary statistics from the posterior predictive distribution of log finish times and the distribution of observed log finish times for three randomly chosen races for both men (top) and women (bottom).}
	\label{tab:pred}
\end{table}

Again, we observe no systematic discrepancies between the summary statistics resulting from the sample from the model and the observations. This is particularly true for the quartiles, although even the minimum and maximum finish times for the races are sometimes more extreme in the predictions and sometimes more extreme in the observations.

\section{Discussion}
\label{sec:discuss}

In this paper we have considered the modelling of cross country finish times for amateur athletes in the NEHL in the North East of England. We conducted an exploratory data analysis which indicated potential differences in the distributions of finish times between different courses and a potential positive effect of distance on finish times, although the inter-individual variability in finish times made inferring relationships difficult. It was even more difficult to identify any potential relationship between either windspeed or rainfall with finish times.

We fitted a linear mixed effects model using Bayesian inference via MCMC, with random athlete, course and season effects and fixed effects of distance, windspeed and rainfall, both in the month of the race and the previous month. Windspeed was found to have no effect and was removed. While this contrasts with the results in \cite{Kne19} who found a negative association between windspeed and performance in the Boston marathon, this could potentially be accounted for by the windspeed measurements used being from a central weather station which could be a number of miles from the race venue. Boston is also a point to point course, and so a headwind would be a headwind for the whole race, whereas in the NEHL athletes run a looped course, and so would experience both a headwind and a tailwind on each lap. 

We found highly variable athlete effects indicating the wide range of individuals taking part in the NEHL. There was more variation in the tails for the female runners, indicting that there were a small number of very slow and very fast women compared to the rest of the runners. The course effects were varied, with a similar ordering for the men and women. Herrington and Thornley were the hardest courses, with Druridge Bay and Gosforth being the easiest. We note that this contrasts with the distributions of raw finish times, which essentially ordered the courses by length. These findings are consistent with the informally gathered qualitative opinions expressed by runners in the NEHL.

Distance, rainfall in the month of the race and rainfall in the previous month were found to have positive effects on finish times, in decreasing order of strength. This is consistent with findings in \cite{Kne19,Vim10,Ber19} for road races. In cross country rainfall leading up to the event is particularly important, as it can cause the course to become muddy and slippery, making running much more difficult and physically demanding.

We found quite different effects in the five seasons of the NEHL considered. There was no obvious pre- to post-Covid effect. Again the effects were consistent between male and female runners. Possible reasons for the season to season variability could be that the condition of the courses and the influence of weather are not fully captured by our rainfall variables or that that the overall populations of runners were systematically changing in terms of their fitness in different seasons (e.g., if the placement of races were close to other important races, meaning runners were tired).

Model adequacy was assessed using in-sample validation - comparing the overall distribution and summary statistics from the model's posterior predictive distribution to the observed distribution of finish times. We found good agreement between samples from the posterior predictive distribution and the data for both men and women, indicating good model fit. While this is sufficient for our purposes of inference, if we were interested in using the model for predictive purposes for individual runners we would need to also undertake out-of-sample validation, for example using the results from the 23/24 NEHL season.

This paper is the first to our knowledge that uses statistical models to make inferences about some of the most important factors influencing finish times of cross country runners. There are some extensions to this work that are worth considering. One to use the inferences from this paper to propose a predictive model for amateur cross country finish times. Another is to extend the model to include more weather and condition covariates such as temperature and snowfall (not uncommon at the NEHL). For example there may be an interaction between rainfall and temperature, as frozen ground may change the effect of rainfall on finish times. Additional covariates such as elevation could improve predictive performance, but were not considered for this work since they affect the difficulty of the course. There was an implicit assumption that all runners were remaining of a similar ability in the data. This is likely to be reasonable over a 5-year period, but if further years of data were used it would be sensible to have some discounting of older results in the athlete effects, as ageing runners are likely to get slower over time and younger runners will typically get faster. This would also help to allow for other factors affecting the ability of individual runners, such as female runners returning after giving birth.

\bibliography{refs}

\begin{thebibliography}{12}
\providecommand{\natexlab}[1]{#1}
\providecommand{\url}[1]{\texttt{#1}}
\expandafter\ifx\csname urlstyle\endcsname\relax
  \providecommand{\doi}[1]{doi: #1}\else
  \providecommand{\doi}{doi: \begingroup \urlstyle{rm}\Url}\fi

\bibitem[Alvero-Cruz et~al.(2020)Alvero-Cruz, Carnero, García, Alacid,
  Correas-Gómez, Rosemann, Nikolaidis, and Knechtle]{Alv20}
José~Ramón Alvero-Cruz, Elvis~A. Carnero, Manuel Avelino~Giráldez García,
  Fernando Alacid, Lorena Correas-Gómez, Thomas Rosemann, Pantelis~T.
  Nikolaidis, and Beat Knechtle.
\newblock Predictive performance models in long-distance runners: A narrative
  review.
\newblock \emph{International Journal of Environmental Research and Public
  Health}, 17\penalty0 (21), 2020.

\bibitem[Arderiu and de~Fondeville(2022)]{Ard22}
Andreu Arderiu and Raphaël de~Fondeville.
\newblock Influence of advanced footwear technology on sub-2 hour marathon and
  other top running performances.
\newblock \emph{Journal of Quantitative Analysis in Sports}, 18\penalty0
  (1):\penalty0 73--86, 2022.

\bibitem[Berke(2019)]{Ber19}
David Berke.
\newblock Performance comparison of long-distance running competitions in
  different meteorology and environment based influential conditions.
\newblock \emph{Idojaras}, 123:\penalty0 313--328, 2019.

\bibitem[Ely et~al.(2007)Ely, Cheuvront, Roberts, and Montain]{Ely07}
Matthew~R Ely, Samuel~N Cheuvront, William~O Roberts, and Scott~J Montain.
\newblock Impact of weather on marathon-running performance.
\newblock \emph{Medicine and science in sports and exercise}, 39\penalty0
  (3):\penalty0 487—493, 2007.

\bibitem[Fogliato et~al.(2021)Fogliato, Oliveira, and Yurko]{Fog21}
Riccardo Fogliato, Natalia~L. Oliveira, and Ronald Yurko.
\newblock Trap: a predictive framework for the assessment of performance in
  trail running.
\newblock \emph{Journal of Quantitative Analysis in Sports}, 17\penalty0
  (2):\penalty0 129--143, 2021.

\bibitem[Keogh et~al.(2019)Keogh, Smyth, Caulfield, Lawlor, Berndsen, and
  Doherty]{Keo19}
Alison Keogh, Barry Smyth, Brian Caulfield, Aonghus Lawlor, Jakim Berndsen, and
  Cailbhe Doherty.
\newblock Prediction equations for marathon performance: A systematic review.
\newblock \emph{International Journal of Sports Physiology and Performance},
  14\penalty0 (9):\penalty0 1159 -- 1169, 2019.

\bibitem[Knechtle et~al.(2019)Knechtle, Di~Gangi, Rüst, Villiger, Rosemann,
  and Nikolaidis]{Kne19}
Beat Knechtle, Stefania Di~Gangi, Christoph~Alexander Rüst, Elias Villiger,
  Thomas Rosemann, and Pantelis~Theo Nikolaidis.
\newblock The role of weather conditions on running performance in the boston
  marathon from 1972 to 2018.
\newblock \emph{PLOS ONE}, 14:\penalty0 1--16, 2019.

\bibitem[Plummer(2022)]{Plu22}
Martyn Plummer.
\newblock \emph{rjags: Bayesian Graphical Models using MCMC}, 2022.
\newblock URL \url{https://CRAN.R-project.org/package=rjags}.
\newblock R package version 4-13.

\bibitem[Stival et~al.(2023)Stival, Bernardi, Cattelan, and Dellaportas]{Sti23}
Mattia Stival, Mauro Bernardi, Manuela Cattelan, and Petros Dellaportas.
\newblock {Missing data patterns in runners’ careers: do they matter?}
\newblock \emph{Journal of the Royal Statistical Society Series C: Applied
  Statistics}, 72\penalty0 (1):\penalty0 213--230, 2023.

\bibitem[Suping et~al.(1992)Suping, Guanglin, Yanwen, and Ji]{Zha92}
Zhang Suping, Meng Guanglin, Wang Yanwen, and Li~Ji.
\newblock Study of the relationships between weather conditions and the
  marathon race, and of meteorotropic effects on distance runners.
\newblock \emph{Int J Biometeorol}, 36:\penalty0 63--68, 1992.

\bibitem[Vihma(2010)]{Vim10}
T~Vihma.
\newblock Effects of weather on the performance of marathon runners.
\newblock \emph{Int J Biometeorol}, 54:\penalty0 297--306, 2010.

\bibitem[Weiss et~al.(2022)Weiss, Valero, Villiger, Thuany, Cuk, Scheer, and
  Knechtle]{Wei22}
Katja Weiss, David Valero, Elias Villiger, Mabliny Thuany, Ivan Cuk, Volker
  Scheer, and Beat Knechtle.
\newblock Relationship between running performance and weather in elite
  marathoners competing in the new york city marathon.
\newblock \emph{Scientific Reports}, 12, 2022.

\end{thebibliography}
\bibliographystyle{plainnat}

\pagebreak

\section*{Appendix}

\begin{figure}[ht]
\centering
\includegraphics[width=0.9\linewidth]{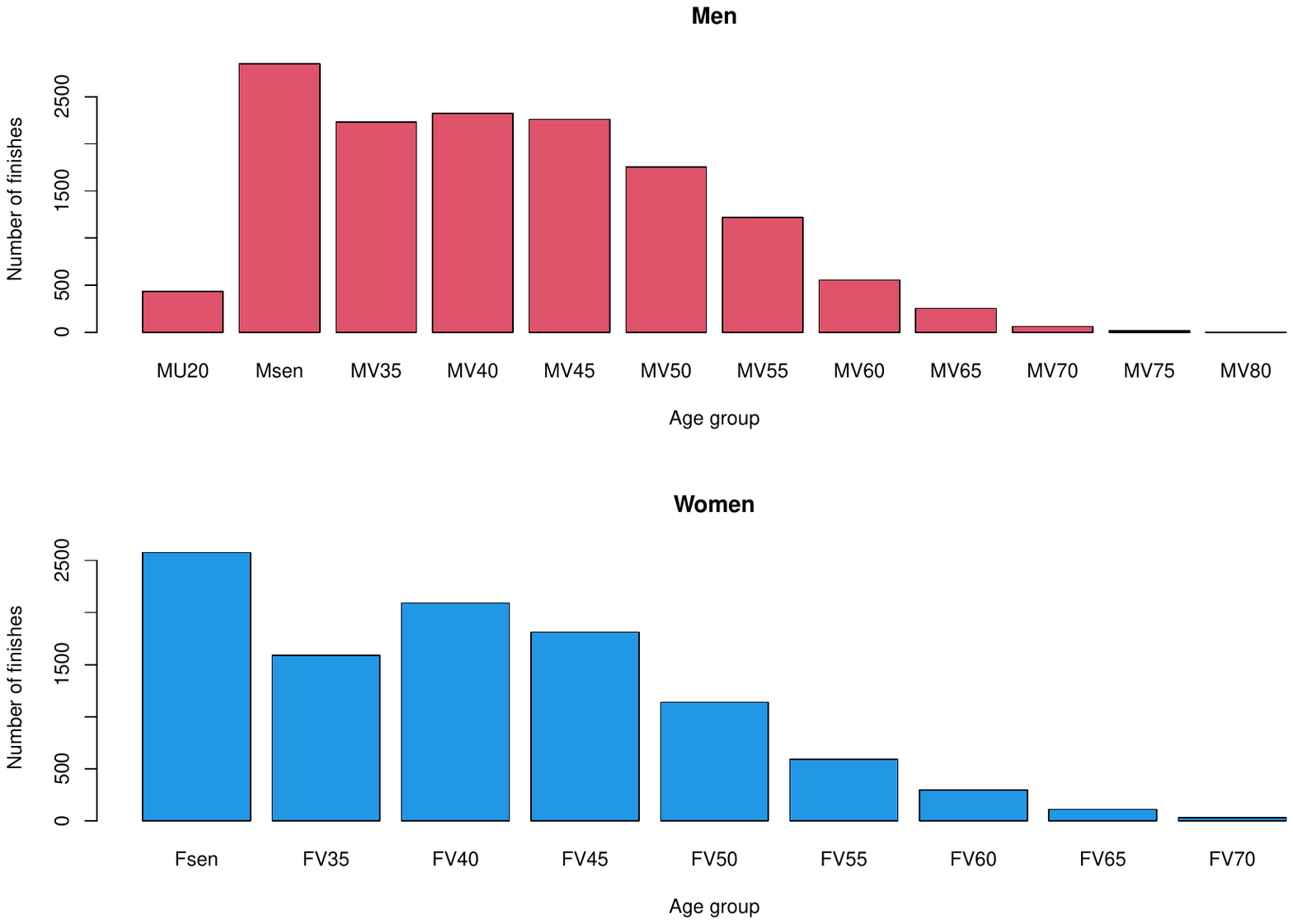}
\caption{The number of finishers by age group across all races for men (top, red) and women (bottom, blue). MU20 represents men 18-20 years old, Msen represents men 20-35, MV35 represents men 36-40, etc. The convention is the same for women, with F replacing M. Women aged 18-20 run in a separate race.}
\label{fig:age}
\end{figure}

\begin{figure}[ht]
	\centering
	\includegraphics[width=0.9\linewidth]{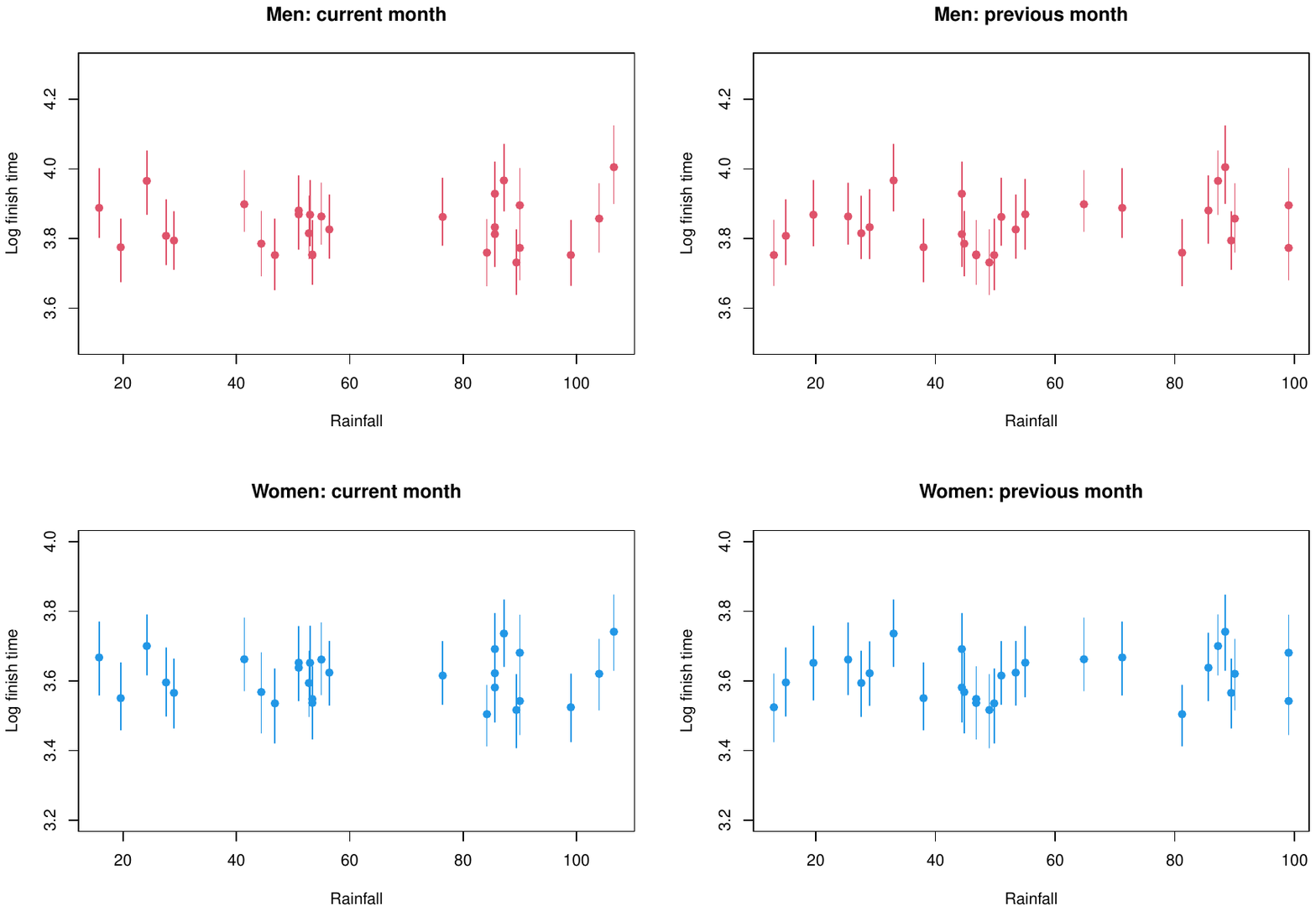}
	\caption{The median and inter-quartile range of the log finish times for each race plotted against the rainfall in the current month (left) and rainfall in the previous month (right) for men (top, red) and women (bottom, blue).}
	\label{fig:rain}
\end{figure}

\begin{table}[ht]
\centering
\begin{tabular}{|c||ccc||ccc|} \hline
	& \multicolumn{3}{|c||}{Men} & \multicolumn{3}{c|}{Women} \\ \hline
Effect & LQ & Med & UQ & LQ & Med & UQ \\ \hline \hline
Alnwick &  0.000 & 0.000 & 0.000 & 0.000 & 0.000 & 0.000 \\
Aykley Heads &  0.016 & 0.017 & 0.019 & 0.029 & 0.031 & 0.033 \\
Druridge Bay &  -0.070 & -0.068 & -0.066 & -0.059 & -0.057 & -0.055 \\
Gosforth & -0.045 & -0.043 & -0.041 & -0.032 & -0.029 & -0.026 \\
Herrington &  0.147 & 0.150 & 0.154 & 0.187 & 0.191 & 0.195 \\
Lambton & -0.005 & -0.003 & -0.002 &  0.000 &  0.002 &  0.003 \\
Thornley Hall &  0.149 & 0.153 & 0.156 & 0.176 & 0.181 & 0.185 \\
Wrekenton & -0.010 & -0.006 & -0.003 & 0.014 & 0.018 & 0.021 \\ \hline \hline
Season 17/18 &  0.000 & 0.000 & 0.000 & 0.000 & 0.000 & 0.000 \\
Season 18/19 &  0.002 & 0.004 & 0.005 & -0.002 & 0.000 & 0.002 \\
Season 19/20 & 0.031 &  0.032 & 0.034 & 0.007 & 0.009 & 0.011 \\
Season 21/22 &  0.044 & 0.045 & 0.046 & 0.026 & 0.028 & 0.029 \\
Season 22/23 & -0.001 & 0.000 & 0.002 &-0.010 & -0.008 & -0.007 \\ \hline \hline
Distance &    0.219 &  0.224 &  0.228 & 0.359 & 0.368 & 0.375 \\
Windspeed &  -0.001 & 0.000 & 0.000 & 0.000 & 0.000 & 0.000 \\
Current rainfall &      0.001 & 0.001 & 0.001 & 0.001 & 0.001 & 0.001 \\
Previous rainfall &   0.001 &  0.001 & 0.001 & 0.001 & 0.001 & 0.001 \\ \hline
\end{tabular}
\caption{The lower quartile (LQ), median (Med) and upper quartile (UQ) of the marginal posterior distribution for each fixed and random effect for the linear mixed effects model including windspeed, for both men and women.}
\label{tab:wind}
\end{table}

\begin{table}[ht]
\centering
\begin{tabular}{|c||ccc||ccc|} \hline
	& \multicolumn{3}{|c||}{Men} & \multicolumn{3}{c|}{Women} \\ \hline
Effect & LQ & Med & UQ & LQ & Med & UQ \\ \hline \hline
Alnwick & 0.000 & 0.000 & 0.000 & 0.000 & 0.000 & 0.000 \\
Aykley Heads & 0.018 & 0.020 & 0.021 & 0.018 & 0.020 & 0.021 \\
Druridge Bay & -0.065 & -0.064 & -0.062 & -0.065 & -0.064 & -0.062 \\
Gosforth & -0.043 & -0.040 & -0.037 & -0.043 & -0.040 & -0.037 \\
Herrington & 0.153 & 0.158 & 0.162 & 0.153 & 0.158 & 0.162 \\
Lambton & -0.001 & 0.000 & 0.002 & -0.001 & 0.000 & 0.002 \\
Thornley Hall & 0.155 & 0.160 & 0.165 & 0.155 & 0.160 & 0.165 \\
Wrekenton & -0.001 & 0.004 & 0.008 & -0.001 & 0.004 & 0.008 \\ \hline \hline
Season 17/18 & 0.000 & 0.000 & 0.000 & 0.000 & 0.000 & 0.000 \\
Season 18/19 & -0.001 & 0.000 & 0.002 & -0.001 & 0.000 & 0.002 \\
Season 19/20 & 0.030 & 0.031 & 0.033 & 0.030 & 0.031 & 0.033 \\
Season 21/22 & 0.042 & 0.043 & 0.045 & 0.042 & 0.043 & 0.045 \\
Season 22/23 & 0.000 & 0.001 & 0.002 & 0.000 & 0.001 & 0.002 \\ \hline \hline
Distance & 0.063 & 0.069 & 0.075 & 0.063 & 0.069 & 0.075 \\
Current rainfall & 0.001 & 0.001 & 0.001 & 0.001 & 0.001 & 0.001 \\
Previous rainfall & 0.001 & 0.001 & 0.001 & 0.001 & 0.001 & 0.001 \\ \hline
\end{tabular}
\caption{The lower quartile (LQ), median (Med) and upper quartile (UQ) of the marginal posterior distribution for each fixed and random effect for the linear mixed effects model with response variable log pace, for both men and women.}
\label{tab:pace}
\end{table}

\begin{figure}[ht]
	\centering
	\includegraphics[width=\linewidth]{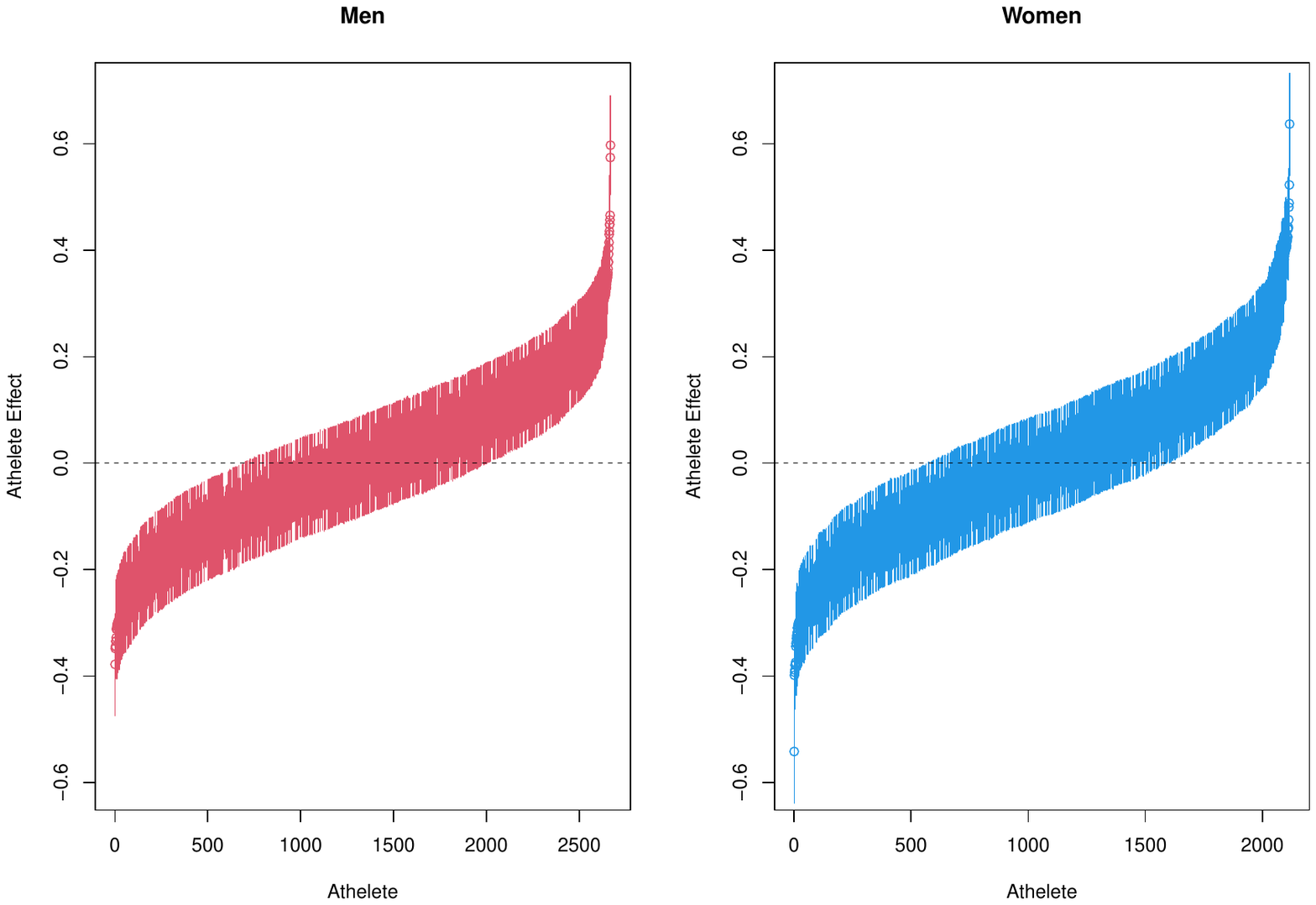}
	\caption{The posterior mean and 95\% posterior symmetric intervals for the athlete effects for men (red, left) and women (blue, right).}
	\label{fig:ath}
\end{figure}

\begin{figure}[ht]
	\centering
	\includegraphics[width=\linewidth]{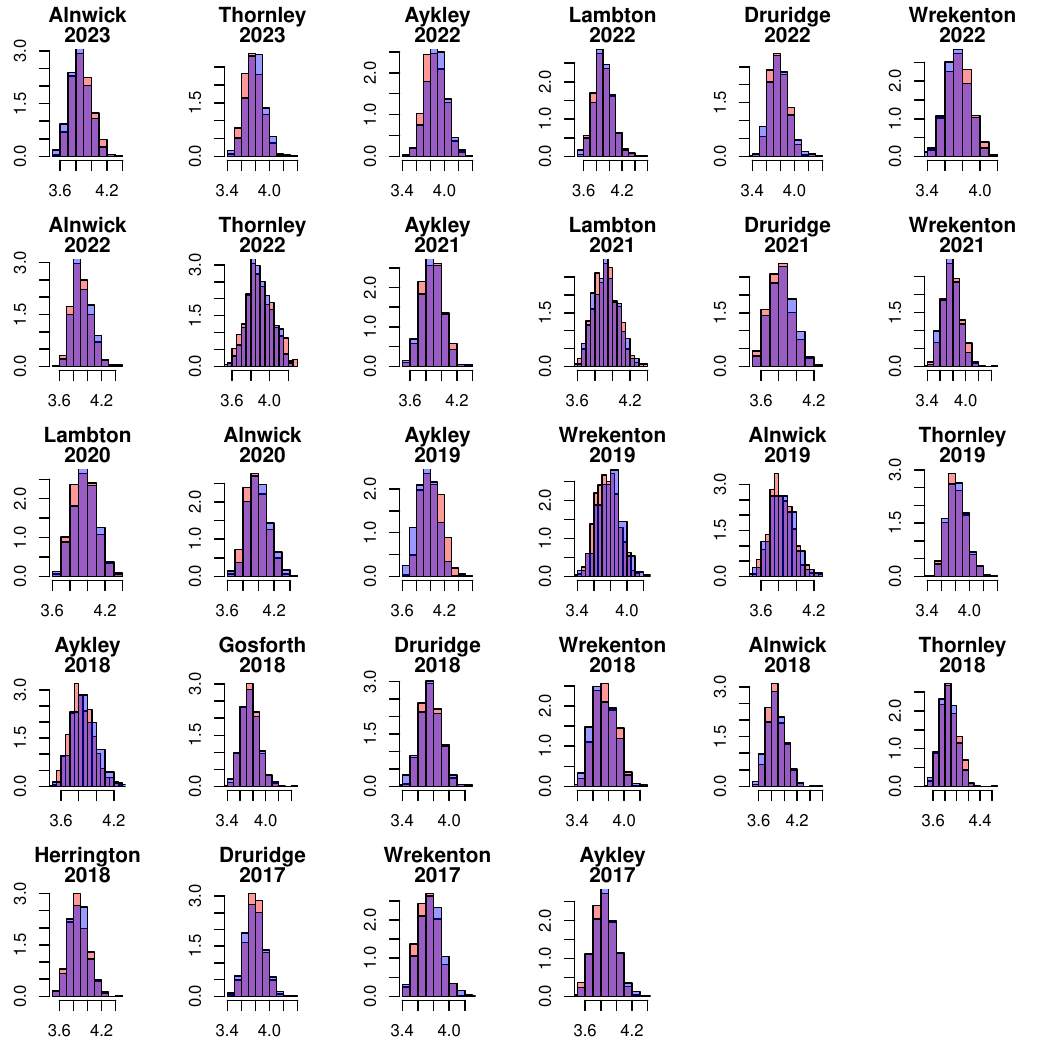}
	\caption{Histograms of the distribution of men's log finish times from a sample from the posterior predictive distribution (blue) and the observed log finish times (red) for each of the 28 races used to fit the model.}
	\label{fig:predmen}
\end{figure}

\begin{figure}[ht]
	\centering
	\includegraphics[width=\linewidth]{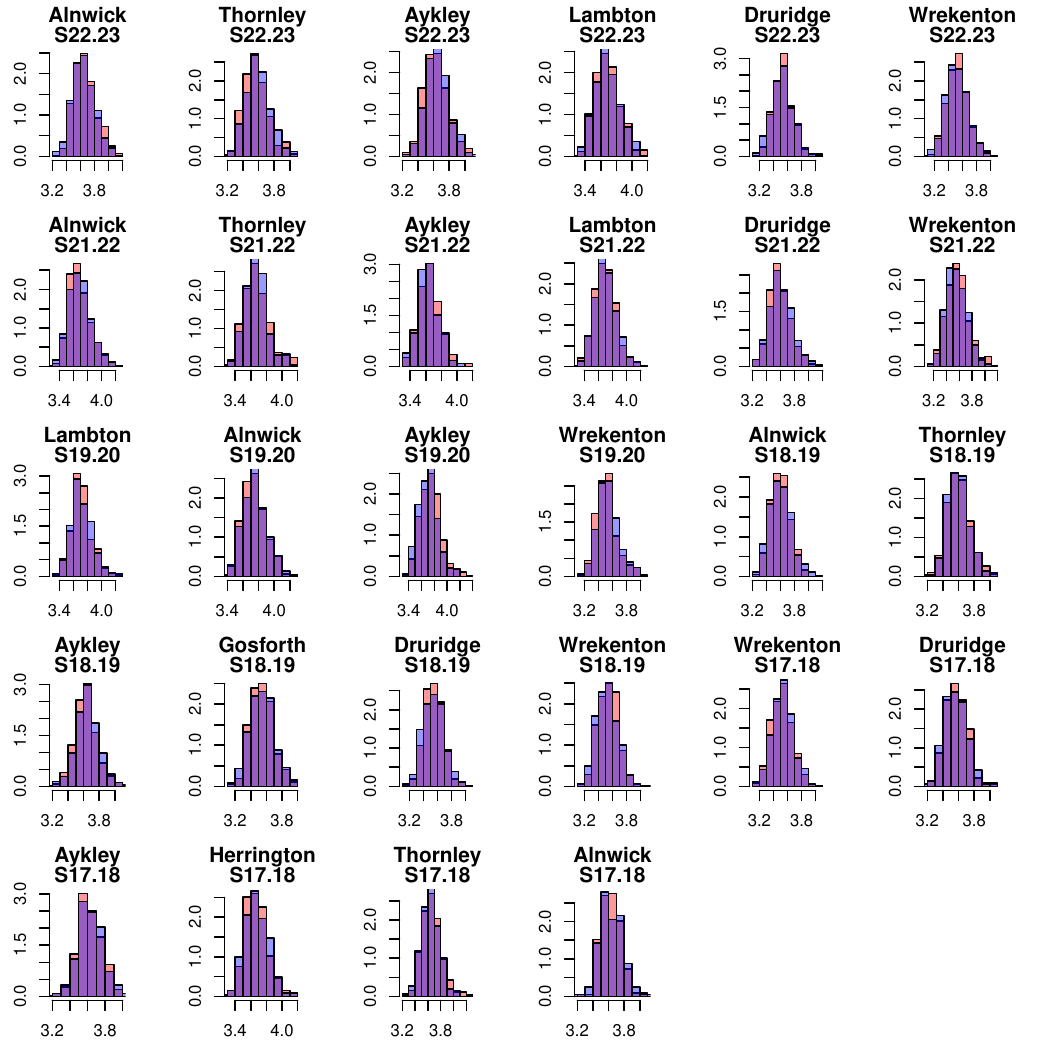}
	\caption{Histograms of the distribution of women's log finish times from a sample from the posterior predictive distribution (blue) and the observed log finish times (red) for each of the 28 races used to fit the model.}
	\label{fig:predwomen}
\end{figure}

\end{document}